\documentclass[letterpaper,9pt,twocolumn,twoside,]{pinp}

\usepackage[T1]{fontenc}
\usepackage[utf8]{inputenc}

\definecolor{pinpblue}{HTML}{185FAF}  
\definecolor{pnasbluetext}{RGB}{101,0,0} %

\usepackage{xtab,booktabs}
\usepackage{xifthen}
\DeclareFontFamily{OT1}{pzc}{}
\DeclareFontShape{OT1}{pzc}{m}{it}{<-> s * [1.250] pzcmi7t}{}
\DeclareMathAlphabet{\mathscr}{OT1}{pzc}{m}{it}
\usepackage{booktabs}
\usepackage{longtable}
\usepackage{array}
\usepackage{multirow}
\usepackage{wrapfig}
\usepackage{float}
\usepackage{colortbl}
\usepackage{pdflscape}
\usepackage{tabu}
\usepackage{threeparttable}
\usepackage{threeparttablex}
\usepackage[normalem]{ulem}
\usepackage{makecell}
\usepackage{xcolor}

\title{Retrospective analysis of a fatal dose-finding trial}

\author[a]{David C. Norris}

  \affil[a]{Precision Methodologies, LLC, Seattle WA, 98102}

\leadauthor{Norris, DC}


\begin{abstract}
The commonplace description of phase 1 clinical trials in oncology as
``primarily concerned with safety'' is belied by their near universal
adoption of dose-escalation practices which are inherently unsafe. In
contrast with dose \emph{titration}, cohort-wise dose \emph{escalation}
regards patients as exchangeable, an indefensible assumption in the face
of widely appreciated inter-individual heterogeneity in pharmacokinetics
and pharmacodynamics (PKPD). I have previously advanced this argument in
terms of a \emph{precautionary coherence} principle that brings the
well-known coherence notion of Cheung (2005) into contact with modern
imperatives of patient-centeredness and precision dosing. Here, however,
I explore these matters in some mechanistic detail by analyzing a trial
of the bispecific T cell engager AFM11, in which a fatal toxicity
occurred. To this end, I develop a Bayesian dose-response model for a
single ordinal toxicity. By constructing this model's priors to align
with the AFM11 trial \emph{as designed and conducted}, I demonstrate the
incompatibility of that design with any reasonable expectation of
safety. Indeed, the model readily yields prospective estimates of toxic
response probabilities that suggest the fatality in this trial could
have been foreseen as likely.
\end{abstract}

\dates{WORKING PAPER DRAFT compiled \today.}

\doifooter{\url{https://cran.r-project.org/package=DTAT}}

\pinpfootercontents{Analysis of a fatal dose-finding trial}

\begin{document}

\verticaladjustment{-2pt}

\maketitle
\thispagestyle{firststyle}
\ifthenelse{\boolean{shortarticle}}{\ifthenelse{\boolean{singlecolumn}}{\abscontentformatted}{\abscontent}}{}


\newcommand{\MTD}[2][]{\ensuremath{\ifthenelse{\isempty{#1}}{\mathrm{MTD}_{#2}}{\mathrm{MTD}_{#1}^{#2}}}}
\newcommand{\Figure}[1]{\mbox{Figure \ref{#1}}}
\newcommand{\Table}[1]{\mbox{Table \ref{#1}}}

\makeatletter
\def\tagform@#1{\maketag@@@{(\ignorespaces#1\unskip\@@italiccorr)}}
\makeatother

\newcommand{\eqsrange}[2]{Eqs.\ (\ref{#1}--\ref{#2})}

\hypertarget{introduction}{%
\section{Introduction}\label{introduction}}

In October 2018, German biopharmaceutical firm Affimed N.V. announced a
clinical hold on two phase 1 trials of its bispecific T cell engager,
AFM11, after 1 fatal and 2 life-threatening neurological toxicities
occurred in the highest-dose cohorts of these trials
\citep{affimed_affimed_2018}. Six months later, Affimed announced the
termination of AFM11 development \citep{affimed_affimed_2019}.

The occurrence of such toxicities in a dose-escalation trial exemplifies
the \emph{precautionary coherence} (PC) principle
\citep{norris_precautionary_2017,norris_ethical_2019,norris_comment_2020}.
The PC principle is itself a straightforward consequence of
inter-individual heterogeneity in pharmacokinetics and pharmacodynamics
(PKPD). Interestingly, Affimed seems to have undertaken the clinical
development of AFM11 with at least some appreciation for the likelihood
of such heterogeneity, and for the consequent relevance of dose
titration. A December 2015 press release \citep{affimed_affimed_2015}
states AFM11 ``is tunable by dosing adjustment when required.'' And
indeed an acknowledgement of PKPD heterogeneity seems to have informed
both AFM11 trials. The first trial,
\href{https://clinicaltrials.gov/ct2/show/NCT02106091}{NCT02106091}, was
formally titled, ``A Pharmacodynamically-guided, Dose-escalation, Phase
I Study.'' The second trial,
\href{https://clinicaltrials.gov/ct2/show/NCT02848911}{NCT02848911}
employed \emph{step-up dosing} within each dose cohort---an accomodation
that virtually acknowledges the PC principle.

\hypertarget{afm11-toxic-dose-response}{%
\section{AFM11 toxic dose-response}\label{afm11-toxic-dose-response}}

The former trial, in CD-19 positive B-cell non-Hodgkin lymphoma (NHL),
has to my knowledge not been reported at a scientific meeting or in the
literature. But the latter trial, in adults with precursor B-cell acute
lymphoblastic leukemia (B-ALL), was presented at the ASH 2018 Annual
Meeting \citep{salogub_phase_2018} in sufficient detail to reveal its
design, dose levels, and the toxic responses for the 17 patients
enrolled.

\begin{table}[!h]

\caption{\label{tab:abstraction}\label{tbl:AFM11-ALL}Doses received and concomitant toxic responses for the 17 patients in AFM11 trial NCT02848911, abstracted from Salogub et al., [2018]. The step-up dosing in this trial enables abstraction of both a non-toxic dose and a maximum dose for each patient. (For the 11 patients who experienced no toxicity at the higher dose, these doses are identical.)}
\centering
\begin{tabular}[t]{>{\raggedleft\arraybackslash}p{0.75in}>{\raggedleft\arraybackslash}p{0.65in}>{\raggedleft\arraybackslash}p{0.65in}>{\raggedleft\arraybackslash}p{0.65in}}
\toprule
Patient & Non-toxic dose
(ng/kg/week) & Highest dose
(ng/kg/week) & CTCAE Grade at highest dose\\
\midrule
\addlinespace[0.3em]
\multicolumn{4}{l}{\textbf{Cohort 1}}\\
\hspace{1em}1 & 0.7 & 2 & 2\\
\hspace{1em}2 & 2 & 2 & 0\\
\addlinespace[0.3em]
\multicolumn{4}{l}{\textbf{Cohort 2}}\\
\hspace{1em}3 & 6 & 6 & 0\\
\addlinespace[0.3em]
\multicolumn{4}{l}{\textbf{Cohort 3}}\\
\hspace{1em}4 & 7 & 20 & 2\\
\hspace{1em}5 & 20 & 20 & 0\\
\hspace{1em}6 & 20 & 20 & 0\\
\addlinespace[0.3em]
\multicolumn{4}{l}{\textbf{Cohort 4}}\\
\hspace{1em}7 & 60 & 60 & 0\\
\hspace{1em}8 & 20 & 60 & 1\\
\hspace{1em}9 & 60 & 60 & 0\\
\addlinespace[0.3em]
\multicolumn{4}{l}{\textbf{Cohort 5}}\\
\hspace{1em}10 & 180 & 180 & 0\\
\hspace{1em}11 & 180 & 180 & 0\\
\hspace{1em}12 & 60 & 180 & 3\\
\hspace{1em}13 & 180 & 180 & 0\\
\hspace{1em}14 & 180 & 180 & 0\\
\addlinespace[0.3em]
\multicolumn{4}{l}{\textbf{Cohort 6}}\\
\hspace{1em}15 & 400 & 400 & 0\\
\hspace{1em}16 & 60 & 130 & 3\\
\hspace{1em}17 & 0 & 130 & 5\\
\bottomrule
\end{tabular}
\end{table}

These doses and toxicities are abstracted in
\mbox{Table \ref{tbl:AFM11-ALL}}. The first 2 cohorts were abbreviated
by an initial accelerated titration \citep{simon_accelerated_1997} that
preceded the 3 + 3 stage of the study. The step-up dosing employed in
this trial, under which each cohort's first week of treatment was
administered at 1/3 of the cohort's target dose, resulted in observation
of some patients (1, 4, 8 and 12) both at a low dose that caused no
toxicity, and at a higher dose that produced a toxicity of CTCAE Grade 1
or higher. Patients 16 and 17 experienced toxicities during the first
week of treatment, so these register at the Cohort 6 starting dose of
130 ng/kg weekly. Patient 16 recovered with a dosing interruption, and
``restarted on Cohort 5 dose'' \citep{salogub_phase_2018}, which I have
taken to mean the Cohort 5 \emph{starting dose} of 60 ng/kg weekly,
rather than the \emph{full} target dose of 180
ng/kg.\footnote{This target dose would have exceeded by 38\% the 130 ng/kg weekly that had caused this patient a Grade 3 toxicity after 3 days of treatment. It seems implausible that titration so far above a patient's previously toxic dose would have been attempted without \citet{salogub_phase_2018} remarking on it.}
Patient 17 died within the first week of treatment, and so never
received a nonzero dose that was demonstrably non-toxic. The non-toxic
dose for this patient is therefore coded as 0. The majority (11/17) of
study participants experienced no toxicity; consequently, their
tabulated non-toxic and highest doses coincide.

\hypertarget{modeling-dose-response-heterogeneity-for-an-ordinal-toxicity}{%
\subsection{Modeling dose-response heterogeneity for an ordinal
toxicity}\label{modeling-dose-response-heterogeneity-for-an-ordinal-toxicity}}

To represent the inter-individual heterogeneity of toxic dose-response,
we suppose that CTCAE toxicity grade \(Y_i \in \{0,1,2,3,4,5\}\) in any
given patient \(i\) is a monotone, left-continuous step function
\(y_i(x)\) of dose \(x\), with the steps occurring at doses
\(\ensuremath{\ifthenelse{\isempty{i}}{\mathrm{MTD}_{g}}{\mathrm{MTD}_{i}^{g}}}, g \in \{0,...,5\}, i \in \{1,...,N\}\):
\begin{equation}\label{eq:defMTDig}
\ensuremath{\ifthenelse{\isempty{i}}{\mathrm{MTD}_{g}}{\mathrm{MTD}_{i}^{g}}} := \max_x \{x | y_i(x) < g \}.
\end{equation} Thus, for example,
\(\ensuremath{\ifthenelse{\isempty{i}}{\mathrm{MTD}_{3}}{\mathrm{MTD}_{i}^{3}}}\)
corresponds to the highest dose that patient \(i\) could tolerate
without experiencing a grade-3 toxicity. The transition from CTCAE
grades 2 to 3 is customarily identified with `intolerability', such that
grade \(\ge 3\) toxicities are (with certain exceptions that won't
concern us here) regarded as `dose-limiting' (DLTs). Thus when \(g=3\)
we may drop the superscript 3 to write `\MTD{i}' in place of
\ensuremath{\ifthenelse{\isempty{i}}{\mathrm{MTD}_{3}}{\mathrm{MTD}_{i}^{3}}},
and refer to this simply as \(i\)'s MTD.

We suppose that \MTD{i} is log-normally distributed in the population:
\begin{equation}
\log \mathrm{MTD}_i \sim \mathscr{N}(\mu, \tau),
\end{equation} where the traditional Bayesian parametrization in terms
of precision \(\tau \equiv \sigma^{-2}\) has been employed.

To promote safer phase 1 trials, our models must support extrapolation
from the more commonly observed toxicity grades \(g \in \{1,...,3\}\) to
the severe and fatal toxicities (\(g \in \{4, 5\}\)) of greatest concern
for participant safety. In the present context, we will enable this by
assuming that our toxicity grading system is somehow aligned with the
underlying pharmacology in a scale-free manner, so that
\((\ensuremath{\ifthenelse{\isempty{i}}{\mathrm{MTD}_{g}}{\mathrm{MTD}_{i}^{g}}})_{g=1}^5\)
is a geometric sequence with ratio \(r_i\): \begin{align}
\mathrm{MTD}_i^g & = r_i^{(g-3)} \cdot \mathrm{MTD}_i \label{eq:MTDig} \\
\log r_i & \sim \mathscr{N}(\log r_0, \tau_r) \label{eq:r_i}
\end{align} While I feel obliged to subscript a quantity such as \(r_i\)
to acknowledge formally its likely inter-individual heterogeneity,
application of this model to the small sample sizes characteristic of
phase 1 trials (and certainly to the 17-patient trial analyzed here)
requires imposing an identifying restriction \(\tau_r \gg 1\). (Had
AFM11's chief toxicity \emph{not} been uniformly neurological in nature
\citep{smith_cancer_2018}, the hope of extrapolating from low to high
toxicity grades via Eqs.\ (\ref{eq:MTDig}--\ref{eq:r_i}) would be much
more dubious.)

Priors for this model are specified as follows, inferred where possible
from Affimed's design of the trial: \begin{align}
\mu & \sim \mathscr{U}(2.9, 7.5) \\
\sqrt{e^{1/\tau}-1} \equiv \mathrm{CV} & \sim \mathscr{N}(0.5, \tau_\mathrm{CV}=36) \\
r_0 & \sim \mathscr{U}(1, 5).
\end{align} The vague uniform prior on \(\mu\) spans an order of
magnitude (2.3 natural logs) either side of a median \MTD{i} at
\(e^{5.2} = 180\) ng/kg weekly, the Cohort 5 target dose. The prior over
\(\tau\) is obtained indirectly from a prior placed on the more
intuitively accessible coefficient of variation (CV) of \MTD{i}. The
\(\tau_\mathrm{CV}=36\) corresponds to \(\sigma=1/6\), which places
\(\mathrm{CV} \ge 1\) in the 3 s.d.~upper tail of the prior. Both the
economics \citep{norris_costing_2017,norris_one-size-fits-all_2018} and
safety \citep{norris_precautionary_2017} of dose-escalation trials
become obviously indefensible for
\(\mathrm{CV}(\ensuremath{\ifthenelse{\isempty{i}}{\mathrm{MTD}_{}}{\mathrm{MTD}_{i}^{}}}) \approx 1\).
Thus, Affimed's choice of a dose-escalation trial suggests it accorded
\(\mathrm{CV}(\ensuremath{\ifthenelse{\isempty{i}}{\mathrm{MTD}_{}}{\mathrm{MTD}_{i}^{}}}) < 1\)
a high prior probability. The vague uniform prior on \(r_0\) extends to
the lowest conceivable limit \(r_0 = 1\), and is centered on \(r_0 = 3\)
in accordance with this trial's threefold separation of dose levels (2,
6, 20, 60, 180, 400) and with the threefold jump chosen for step-up
dosing.

\hypertarget{model-implementation}{%
\subsection{Model implementation}\label{model-implementation}}

The model is implemented in JAGS \citep{plummer_jags:_2003} version
4.3.0, as
below.\footnote{By contrast with the overloaded {\tt I(,)} construct familiar to users of BUGS, the {\tt dinterval} distribution in JAGS is an `observable function' that specifically implements interval censoring. See sections 7.2 and 9.2.4 of the JAGS Version 4.3.0 User Manual (28 June 2017).}

\begin{verbatim}
var N; # number of patients
var okdose[N]; # a tox-free dose for each pt
var aedose[N]; # a toxic dose for each pt

var mtd[N];  # MTDi's on absolute dose scale
var mu, tau; # MTDi ~ dlnorm(mu, tau)
var cv;   # tau <- 1/log(cv^2+1)
var Y[N]; # Toxicity grade for patient i=1,..,N
var Z[N]; # Dummy array of zeros for no-tox obs

data {
  for (i in 1:N) {
    Z[i] <- 0;
  }
}

model {
  mu ~ dunif(2.9, 7.5);
  cv ~ dnorm(0.5, 36); # sigma=1/6 => 1=mean+3sd
  tau <- 1/log(cv^2+1);
  r0 ~ dunif(1, 5);
  for (i in 1:N) {
    mtd[i] ~ dlnorm(mu, tau);
  }
  for (i in 1:N) {
    r[i] ~ dlnorm(log(r0), 50); # so CV(r[i]) ~ 14%
    Y[i] ~ dinterval(aedose[i]/mtd[i],
              c(r[i]^-2, 1/r[i], 1, r[i], r[i]^2));
    Z[i] ~ dinterval(okdose[i]/mtd[i], r[i]^-2);
  }
}
\end{verbatim}

R package \texttt{runjags} \citep{runjags2016} version 2.0.4-6 was used
to run JAGS with 4 chains, each sampled 2500 times with a thinning
interval of 10, following adaptation over 1000 iterations and burn-in of
4000.

\hypertarget{model-estimates}{%
\subsection{Model estimates}\label{model-estimates}}

\mbox{Table \ref{tbl:fit-summary}} summarizes the model fit, showing
potential scale reduction factors near unity and large effective sample
sizes for all parameters, indicative of adequate convergence and
precision.

\begin{table}[!h]

\caption{\label{tab:fitted}\label{tbl:fit-summary}Selected quantiles, mean, and diagnostics for 10,000 MCMC samples from the model of AFM11 toxic dose-response. (SSeff: effective sample size; psrf: potential scale reduction factor.)}
\centering
\begin{tabular}[t]{lrrrrrr}
\toprule
  & Lower95 & Median & Upper95 & Mean & SSeff & psrf\\
\midrule
\addlinespace[0.3em]
\multicolumn{7}{l}{\textbf{Hyperparameters}}\\
\hspace{1em}mu & 4.483 & 5.033 & 5.588 & 5.037 & 6227 & 1.001\\
\hspace{1em}cv & 0.869 & 1.069 & 1.292 & 1.072 & 8515 & 1.001\\
\hspace{1em}tau & 0.980 & 1.312 & 1.710 & 1.333 & 8280 & 1.001\\
\hspace{1em}r0 & 1.168 & 1.328 & 1.539 & 1.336 & 3773 & 1.002\\
\addlinespace[0.3em]
\multicolumn{7}{l}{\textbf{Cohort 1}}\\
\hspace{1em}mtd[1] & 2.000 & 2.641 & 3.490 & 2.698 & 5334 & 1.001\\
\hspace{1em}mtd[2] & 6.818 & 155.074 & 741.512 & 244.976 & 4245 & 1.000\\
\addlinespace[0.3em]
\multicolumn{7}{l}{\textbf{Cohort 2}}\\
\hspace{1em}mtd[3] & 10.616 & 150.651 & 697.634 & 232.181 & 5990 & 1.000\\
\addlinespace[0.3em]
\multicolumn{7}{l}{\textbf{Cohort 3}}\\
\hspace{1em}mtd[4] & 20.000 & 24.622 & 32.626 & 25.318 & 7635 & 1.001\\
\hspace{1em}mtd[5] & 24.685 & 164.743 & 718.587 & 257.397 & 3704 & 1.001\\
\hspace{1em}mtd[6] & 26.416 & 162.084 & 712.577 & 246.507 & 4612 & 1.000\\
\addlinespace[0.3em]
\multicolumn{7}{l}{\textbf{Cohort 4}}\\
\hspace{1em}mtd[7] & 68.506 & 229.548 & 817.905 & 320.609 & 4064 & 1.001\\
\hspace{1em}mtd[8] & 64.879 & 102.858 & 161.341 & 108.212 & 4444 & 1.001\\
\hspace{1em}mtd[9] & 61.216 & 228.808 & 786.405 & 311.936 & 5078 & 1.000\\
\addlinespace[0.3em]
\multicolumn{7}{l}{\textbf{Cohort 5}}\\
\hspace{1em}mtd[10] & 183.516 & 463.337 & 1229.780 & 567.250 & 5316 & 1.001\\
\hspace{1em}mtd[11] & 190.979 & 460.132 & 1237.230 & 566.082 & 5256 & 1.001\\
\hspace{1em}mtd[12] & 134.969 & 159.790 & 179.996 & 158.797 & 10000 & 1.000\\
\hspace{1em}mtd[13] & 196.548 & 462.334 & 1267.920 & 570.773 & 4241 & 1.001\\
\hspace{1em}mtd[14] & 183.836 & 460.362 & 1237.710 & 567.768 & 4652 & 1.002\\
\addlinespace[0.3em]
\multicolumn{7}{l}{\textbf{Cohort 6}}\\
\hspace{1em}mtd[15] & 419.005 & 855.336 & 1938.600 & 988.780 & 5211 & 1.001\\
\hspace{1em}mtd[16] & 106.057 & 119.836 & 129.999 & 119.013 & 10165 & 1.001\\
\hspace{1em}mtd[17] & 11.938 & 50.237 & 95.173 & 52.473 & 9315 & 1.000\\
\bottomrule
\end{tabular}
\end{table}

\begin{figure}

{\centering \includegraphics{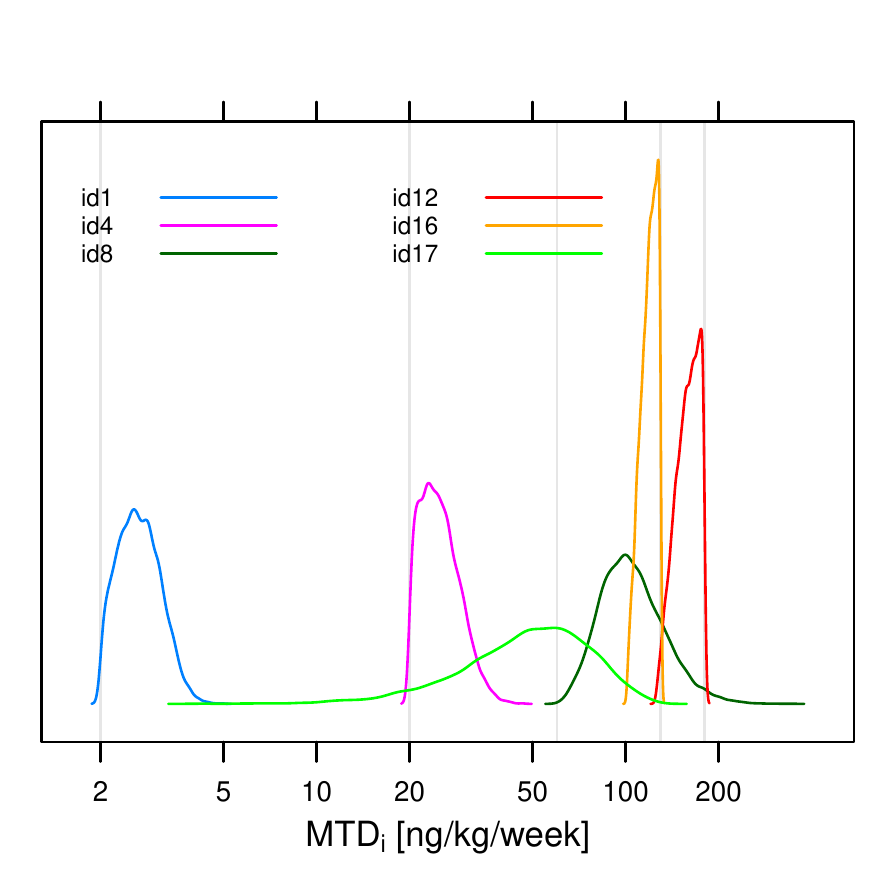} 

}

\caption{\label{fig:tox-densities}Posterior densities of MTD$_i$ (on log scale) for patients who experienced toxicities. The vertical lines show the 5 doses at which the toxicities occurred. Patients 1 and 4, who experienced grade 2 toxicity at dose levels 1 and 3, have sharp lower bounds placed thereby on their MTD$_i$'s as expected. Conversely, patients 12 and 16, who experienced grade 3 toxicities (DLTs), have {\em upper bounds} placed on their MTD$_i$'s. The grade 1 and 5 toxicities experienced by patients 8 and 17 impose somewhat softer bounds due to the slack created by the unknown $r_0$ parameter.}\label{fig:tox-densities}
\end{figure}

\begin{figure}

{\centering \includegraphics{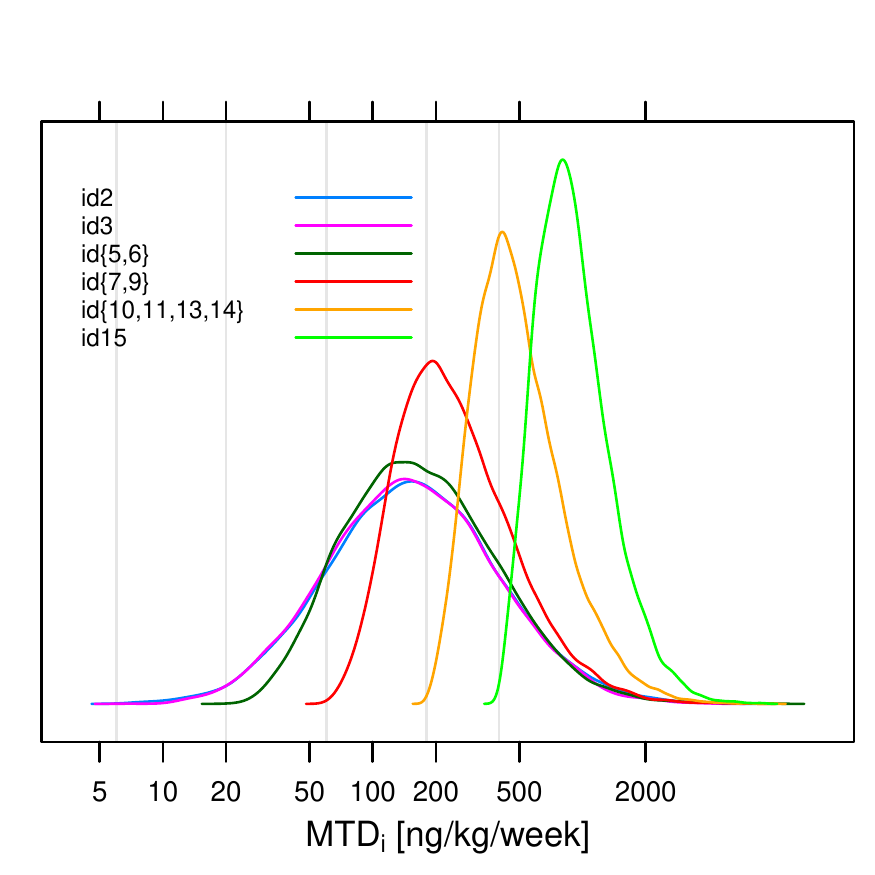} 

}

\caption{\label{fig:toxless-densities}Posterior densities of MTD$_i$ (on log scale) for patients who did not experience toxicities. The vertical lines show the 6 cohort target dose levels. Patients with identical data are of course indistinguishable from the perspective of the model; the samples from such sets of `equivalent' patients have been pooled.}\label{fig:toxless-densities}
\end{figure}

Figures \ref{fig:tox-densities} and \ref{fig:toxless-densities} exhibit
separately the posterior densities of MTD\(_i\) for patients with and
without observed toxicities. The comparison between these Figures
illustrates that observed toxicities constitute the main source of
information in dose-finding trials.

\begin{figure}

{\centering \includegraphics{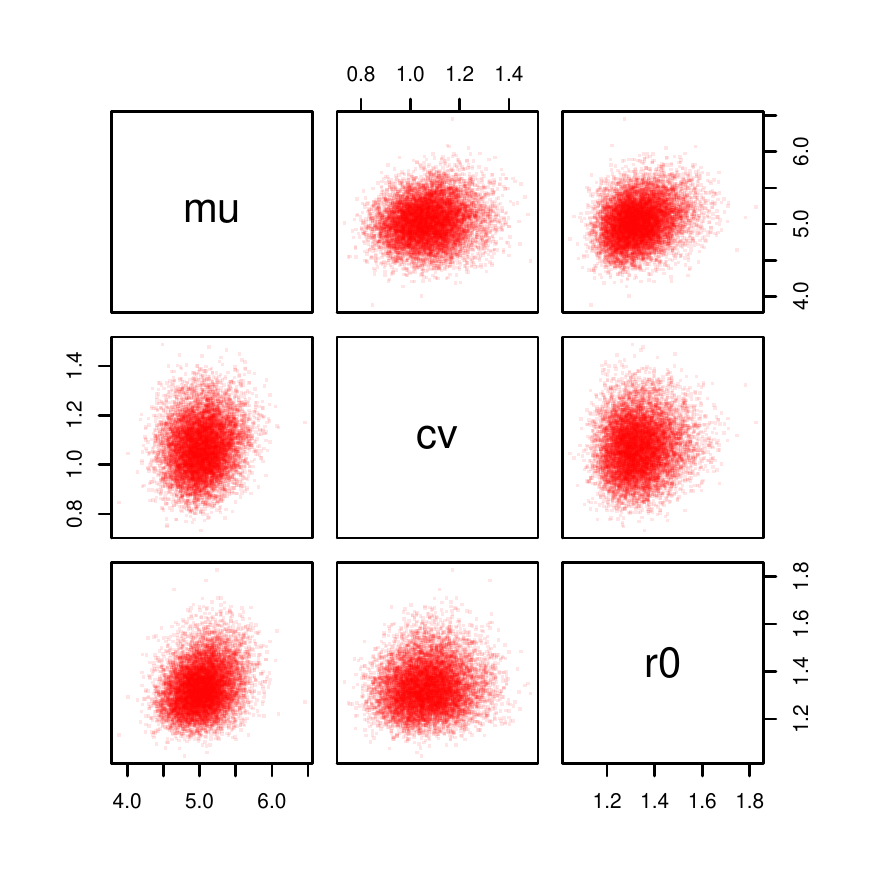} 

}

\caption{\label{fig:hyperparameters}Pairwise joint posterior distributions for model hyperparameters.}\label{fig:examine-fit}
\end{figure}

\mbox{Figure \ref{fig:hyperparameters}} shows the pairwise joint
posterior densities of model hyperparameters \(\mu\), \(\mathrm{CV}\)
and \(r_0\). Notably, the posterior for \(\mathrm{CV}\) is centered
around \(\mathrm{CV} \approx 1\), which we had accorded a low prior
probability. Thus, we are `surprised' to discover the substantial
heterogeneity of toxic dose-response revealed by the severe toxicities
in this trial. Similarly, the posterior density centering
\(r_0 \approx 1.3\), while unremarkable from the standpoint of
pharmacologic intuition, suggests that in retrospect the threefold
dose-level jumps in this trial would have been too large even under
circumstances where \(\mathrm{CV} \ll 1\).

\hypertarget{discussion}{%
\section{Discussion}\label{discussion}}

Given the sparse data in this small dose-escalation study, my analysis
necessarily makes strong assumptions. Possibly the most consequential of
these has been to ignore \emph{exposure time}. It seems likely that
\MTD{i}, far from being constant for each individual, is in fact a
stochastic process,
\(\ensuremath{\ifthenelse{\isempty{i}}{\mathrm{MTD}_{}}{\mathrm{MTD}_{i}^{}}}(t)\).
Treating it here as constant thus ignores time-to-event considerations
(cf.~\citet{cheung_sequential_2000}) which may have been important in
this trial. Was the step-up dosing in this trial conceived as
`conditioning' patients as with `ramp-up' dosing of venetoclax to avoid
tumor lysis syndrome \citep{roberts_targeting_2016}? If so, then
abstracting away time overlooks what may be the crucial pharmacologic
point in AFM11 dosing.

\hypertarget{comparison-with-bekele-thall-2004}{%
\subsection{Comparison with Bekele \& Thall
{[}2004{]}}\label{comparison-with-bekele-thall-2004}}

Formally, Eqs.\ (\ref{eq:defMTDig}--\ref{eq:r_i}) invite comparison with
the ordinal probit model of \citet{bekele_dose-finding_2004}, in which a
latent \(Z\) variable acts like our MTD\(_i\) to determine toxicities
through a relation analogous to
\eqref{eq:defMTDig}:\footnote{Here, and throughout this comparison, I drop the extra $j$ subscript which \citet{bekele_dose-finding_2004} introduced to index multiple toxicities.}
\begin{equation}\label{eq:BT1}
Y = y_k \quad\mbox{if}\quad \gamma_k \le Z < \gamma_{k+1},\; k \in \{0, 1, ..., C\}.
\end{equation} Unlike our MTD\(_i\), however, the latent \(Z\) remains
for \citet{bekele_dose-finding_2004} a unitless abstraction, bereft of a
realistic pharmacologic interpretation. Similarly, they motivate their
log-scaling of dose as a technical maneuver ``{[}t{]}o improve numerical
stability,'' whereas here I appeal to a prior implicit in the
geometrical sequence of pre-specified AFM11 dose levels.

The \((\gamma_k)_{k=0}^{C}\) of \citet{bekele_dose-finding_2004} are
likewise analogous to our
\((\ensuremath{\ifthenelse{\isempty{i}}{\mathrm{MTD}_{g}}{\mathrm{MTD}_{i}^{g}}})_{g=0}^5\).
Because Bekele \& Thall enjoyed access to substantial quantities of
pseudodata via intensive prior elicitation, they placed only vague
priors on their \(\gamma_k\)'s and did not require an identifying
restriction such as \eqref{eq:MTDig}.

\hypertarget{comparison-with-van-meter-et-al.-2012}{%
\subsection{Comparison with Van Meter et
al.~{[}2012{]}}\label{comparison-with-van-meter-et-al.-2012}}

While the formal developments in \citet{van_meter_dose-finding_2012} do
not present close analogues with
Eqs.\ (\ref{eq:defMTDig}--\ref{eq:r_i}), the authors' efforts to ground
their modeling in pharmacologic realism warrant discussion. The `CR-CRM'
model developed by these authors employs a \emph{continuation ratio}
(CR) model, which they introduce in conjunction with an appeal to
pharmacologically realistic intuitions about toxicity:

\begin{quote}
``[T]he CR model allows for comparisons between individuals in a specific toxicity category versus all individuals that experienced a more severe toxicity grade. The CR model also distinguishes between subjects who reached a certain toxicity grade but did not advance to a more severe toxicity and assumes that individuals must `pass through'
the ordinal toxicities to advance to the next highest category. In any clinical trial setting, this assumption is reasonable given that we assume a patient who experiences a grade 3 severe toxicity first presented with symptoms resembling less severe grade 1 or 2 toxicities.''
\end{quote}

It is remarkable to find within the statistical dose-finding literature
this type of inchoate realism, alluding to patients as non-exchangeable
individuals.

The CR model stipulates that, for a toxicity grading system with \(k+1\)
ordinal grades \(\{0, 1, ..., k\}\), the population distribution of
toxic responses is such that \citep[p.304]{van_meter_dose-finding_2012}:
\begin{equation}\label{eq:CR}
\mathrm{logit}[ \mathrm{Pr}(Y=j | Y \ge j, \mathrm{dose}=x) ] = \alpha + \theta_j + \lambda x; j=0,...,k-1,
\end{equation} where it is understood that \(\theta_0 = 0\). Under the
widely adopted CTCAE toxicity grading system, for example, we have
\(k=5\) and the CR model has 6 parameters:
\(\{\alpha, \theta_1, \theta_2, \theta_3, \theta_4, \gamma\}\).

The term `continuation ratio' applies here because the left-hand side of
\eqref{eq:CR} admits interpretation as the log odds for an individual
who has reached a grade-\(j\) toxicity not continuing to a higher grade.
The CR models arise in psychometrics, where they form one class of
polytomous item response theory (IRT) models
\citet{hemker_measurement_2001}. The natural toxicologic analogy would
be that physiology presents a sequence of biochemical barriers to
toxicity not unlike the difficulties presented by a sequentially scored
test item. The grade of toxicity produced by a toxin is then determined
by how far it penetrates through these sequential barriers, much as the
item score achieved by a test-taker depends on how far he advances
through the stages of item difficulty. Whereas psychometrics readily
posits latent variables with realistic interpretations, however, the
statistical literature on dose finding has thus far proven impervious to
such thinking \citep{norris_comment_2020}. Further development of the
CR-CRM model in terms of latent quantities such as MTD\(_i\) may well
serve to bridge that literature into the modern era of precision
medicine, with its expectations of individualized dosing.

\hypertarget{anticipating-and-averting-the-fatal-toxicity-of-patient-17}{%
\subsection{Anticipating and averting the fatal toxicity of patient
17}\label{anticipating-and-averting-the-fatal-toxicity-of-patient-17}}

Citing ``the ethical need to control the probability of overdosing'' in
dose-escalation trials, \citet{babb_cancer_1998} developed the
\emph{escalation with overdose control} (EWOC) criterion. Under this
scheme, each enrolling patient's dose is chosen so as to limit the
probability of an `overdose', defined as exceeding some population
quantile of the MTD\(_i\)
distribution.\footnote{Working as they did entirely {\em within} the one-size-fits-all dose-finding framework, \citet{babb_cancer_1998} did not explicitly recognize this quantile as such, but referred to it simply as "the MTD."}
In the same spirit, we may ask what the probabilities of each grade of
toxicity \emph{would have been} in Cohort 6, conditional on all
observations up through Cohort 5. Technically, this can be accomplished
by dropping Cohort 6 patients 15--17 from the data set, and replacing
them with 2 hypothetical patients: one (at index \(i=15^*\)) starting
the 130 ng/kg weekly dose, and the second (at \(i=16^*\)) stepping up to
the full 400 ng/kg weekly, after no toxicity at the starting dose. The
unobserved nodes \texttt{Y{[}15{]}} and \texttt{Y{[}16{]}} may then be
sampled to \emph{pre}dict the distribution of toxicities in Cohort
6.\footnote{See section 3.2, page 15 of the JAGS Version 4.3.0 User Manual.}
\mbox{Table \ref{tbl:fit-summary-H}} shows the hyperparameter estimates
without the Cohort 6 data remain reasonably close to those in
\mbox{Table \ref{tbl:fit-summary}}.

\begin{table}[!h]

\caption{\label{tab:fitted-H}\label{tbl:fit-summary-H}Selected quantiles, mean, and diagnostics for 10,000 MCMC hyperparameter samples from the model estimated against data from Cohorts 1--5. Also sampled is MTD$_{15^*}$, by construction a random sample from the posterior distribution of population MTD$_i$ immediately before the first enrollment in Cohort 6.}
\centering
\begin{tabular}[t]{lrrrrrr}
\toprule
  & Lower95 & Median & Upper95 & Mean & SSeff & psrf\\
\midrule
\addlinespace[0.3em]
\multicolumn{7}{l}{\textbf{Hyperparameters}}\\
\hspace{1em}mu & 4.509 & 5.167 & 5.808 & 5.176 & 3272 & 1.0007\\
\hspace{1em}cv & 0.847 & 1.056 & 1.269 & 1.058 & 7508 & 1.0000\\
\hspace{1em}tau & 0.992 & 1.335 & 1.752 & 1.360 & 7854 & 1.0000\\
\hspace{1em}r0 & 1.166 & 1.407 & 1.704 & 1.424 & 1972 & 1.0006\\
\addlinespace[0.3em]
\multicolumn{7}{l}{\textbf{Population MTDi}}\\
\hspace{1em}mtd[15*] & 10.278 & 173.204 & 817.957 & 272.691 & 4295 & 1.0002\\
\bottomrule
\end{tabular}
\end{table}

\begin{figure}

{\centering \includegraphics{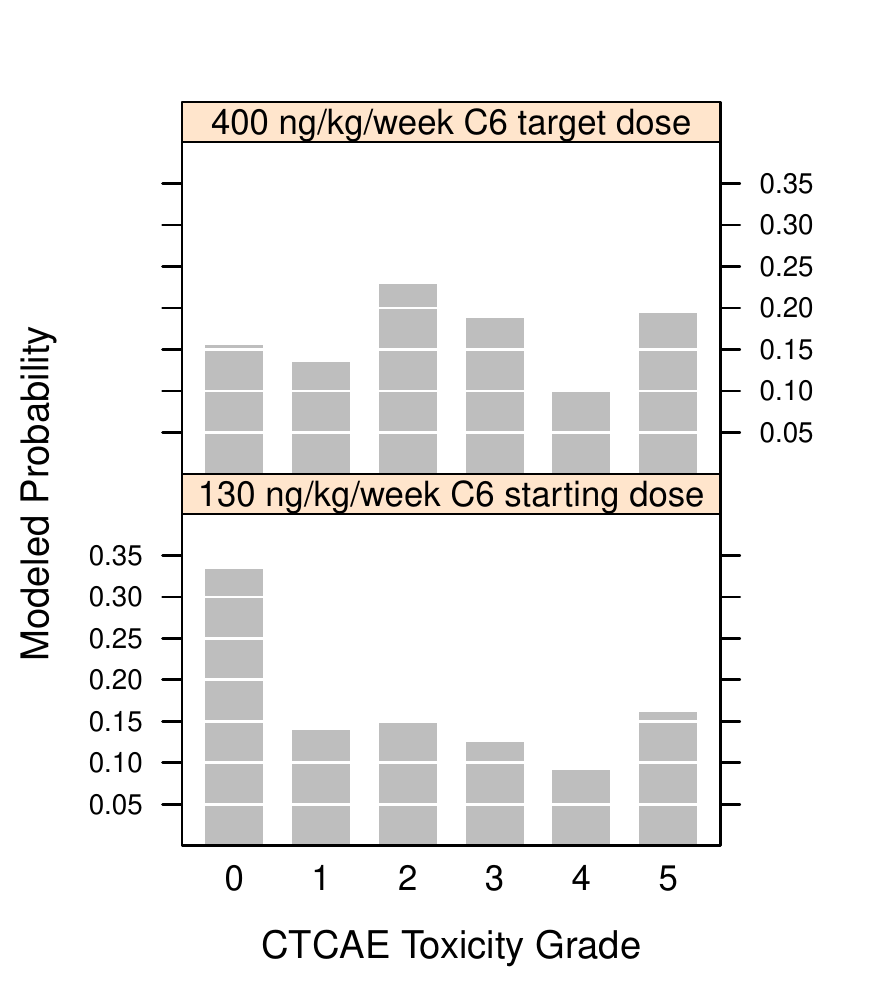} 

}

\caption{\label{fig:fatality-foreseen}Probabilities of Cohort 6 toxicities by CTCAE Grade, as would have been predicted by the model prior to enrollment. Distributions are given separately for the Cohort 6 starting dose, and for step-up to the full target dose conditional on no toxicity with the starting dose. (With 10,000 MCMC samples, the Monte Carlo standard errors are too small to display.)}\label{fig:fatality-foreseen}
\end{figure}

\mbox{Figure \ref{fig:fatality-foreseen}} exhibits the toxicity
distributions for these hypothetical patients. The model is seen to
predict that each patient enrolling in Cohort 6 risks a fatal toxicity
even at the starting dose, with probability exceeding 16\%. More
realistic prior elicitation and modeling of AFM11 toxic dose-response
thus might have informed safer design and conduct of this trial.

\hypertarget{conclusion}{%
\section{Conclusion}\label{conclusion}}

I have developed here a Bayesian model of heterogeneous dose-response
for an ordinal toxicity, and used it to examine retrospectively the
safety of dose-finding trial that was halted for a fatal toxicity. The
model supports an interim analysis suggesting the fatalilty in this
trial could have been anticipated and averted. Without loss of
generality, this mode of analysis may be applied at every dosing
decision in a dose-finding trial. Thus, it forms the basis for designing
and implementing safer dose-finding designs in the future. With access
to sufficient pseudodata from prior elicitation, some of the strong
identifying restrictions employed here may (and should) be relaxed.
Being grounded in realistic pharmacologic intuitions, and explicitly
recognizing patients as non-exchangeable individuals, this model may
prove well suited to such prior elicitation with clinicians.

\hypertarget{data-availability}{%
\subsection{Data availability}\label{data-availability}}

Code for reproducing all of this paper's Figures and analyses may be
found at \href{https://osf.io/9x6j7/}{doi:10.17605/osf.io/9x6j7}.

\hypertarget{competing-interests}{%
\subsection{Competing interests}\label{competing-interests}}

The author operates a scientific and statistical consultancy focused on
precision-medicine methodologies such as those advanced in this article.

\hypertarget{grant-information}{%
\subsection{Grant information}\label{grant-information}}

No grants supported this work.


\bibliography{Affimed,OrdinalTox,Coherence,packages}
\bibliographystyle{plainnat}

\end{document}